\documentclass[12pt]{article}
\usepackage{graphicx}
\usepackage{amsmath}
\usepackage[backend=biber,
bibencoding=utf8, 
hyperref,
style=numeric-comp,
sorting=none,
giveninits=true]{biblatex}
\usepackage{xspace}
\usepackage{hyperref}
\usepackage[english]{babel}
\DeclareBibliographyDriver{unpublished}{%
	\usebibmacro{bibindex}%
	\usebibmacro{begentry}%
	\usebibmacro{author}%
	\setunit{\labelnamepunct}\newblock
	\usebibmacro{title}%
	\newunit
	\printlist{language}%
	\newunit\newblock
	\usebibmacro{byauthor}%
	\newunit\newblock
	\printfield{howpublished}%
	\newunit\newblock
	\printfield{note}%
	\newunit\newblock
	\usebibmacro{location+date}%
	\newunit\newblock
	\usebibmacro{doi+eprint+url}
	\newunit\newblock
	\usebibmacro{addendum+pubstate}%
	\setunit{\bibpagerefpunct}\newblock
	\usebibmacro{pageref}%
	\usebibmacro{finentry}}

\newcommand\pubnumber{CR-2023/006}
\newcommand\pubdate{\today}

\def\institute{Institute of Experimental Physics\\
University of Hamburg, D-22761 Hamburg, GERMANY}
\def\support{\footnote{Work supported by the Ministry for Education and Research (BMBF).\\

Copyright 2023 CERN for the benefit of the ATLAS and CMS Collaborations. Reproduction of this article or parts of it is allowed as specified in the CC-BY-4.0 license
}}

\def\Title#1{\begin{center} {\Large #1 } \end{center}}
\def\Author#1{\begin{center}{ \sc #1} \end{center}}
\def\Address#1{\begin{center}{ \it #1} \end{center}}

\newcommand\pubblock{\rightline{\begin{tabular}{l} \pubnumber\\
         \pubdate  \end{tabular}}}
\newenvironment{Abstract}{\begin{quotation}  }{\end{quotation}}
\newenvironment{Presented}{\begin{quotation} \begin{center} 
             PRESENTED AT\end{center}\bigskip 
      \begin{center}\begin{large}}{\end{large}\end{center} \end{quotation}}
\def\Acknowledgements{\bigskip  \bigskip \begin{center} \begin{large}
             \bf ACKNOWLEDGEMENTS \end{large}\end{center}}

\def\beq{\begin{equation}}
\def\eeq#1{\label{#1}\end{equation}}
\def\eeqn{\end{equation}}

\def\beqa{\begin{eqnarray}}
\def\eeqa#1{\label{#1}\end{eqnarray}}
\def\eeqan{\end{eqnarray}}

\let\bar=\overbar

\def\L{{\cal L}}

\def\O{{\cal O}}

\def\Dslash{\not{\hbox{\kern-4pt $D$}}}
\def\dslash{\not{\hbox{\kern-2pt $\del$}}}

\def\msb{{\bar{\ssstyle M \kern -1pt S}}}

\newcommand{\ttZ}{\ensuremath{\text{t}\bar{\text{t}}\text{Z}}\xspace}
\newcommand{\tZq}{\ensuremath{\text{t}\text{Z}\text{q}}\xspace}
\newcommand{\tZW}{\ensuremath{\text{t}\text{Z}\text{W}}\xspace}
\newcommand{\ttbar}{\ensuremath{\text{t}\bar{\text{t}}}\xspace}

\begin{document}
\begin{titlepage}
\pubblock

\vfill
\Title{Machine Learning in Top Physics in the ATLAS and CMS Collaborations}
\vfill
\Author{ Philip Keicher\support}
\Address{\institute}
\vfill
\begin{Abstract}
Machine learning is essential in many aspects of top-quark related physics in the ATLAS and CMS Collaborations.
This work aims to give a brief overview over current applications in the two collaborations as well as on-going studies for future applications.
\end{Abstract}
\vfill
\begin{Presented}
$15^\mathrm{th}$ International Workshop on Top Quark Physics\\
Durham, UK, 4--9 September, 2022
\end{Presented}
\vfill
\end{titlepage}
\def\thefootnote{\fnsymbol{footnote}}
\setcounter{footnote}{0}

\section{Introduction}

Machine learning~(ML) applications are nowadays well-established tools in modern data science problems.
An increasing number of techniques and analyses in high-energy physics are harnessing the capabilities of these methods.
Many of these applications within the community are related to the top quark due to its rich phenomenology within both the standard model of particle physics~(SM) and beyond.
Most of these applications for top quark physics within the ATLAS~\cite{det:ATLAS} and CMS~\cite{det:CMS} Collaborations can be separated according to three paradigms.
First, ML applications are used as tagging algorithms to identify top quarks.
Second, ML is used to reconstruct top quark properties.
Finally, ML is used to construct observables that have a strong separation of signal and background processes, thus further increasing the statistical power of analyses.
Some select examples of the usage of ML techniques in these fields within both ATLAS and CMS Collaborations can be found in the references of this work.

This work focuses on the last field of applications described above.
Specifically, the ML usage of an example from each collaboration and a summary of select ongoing projects is presented.
For a more thorough review of the analyses, please refer to Refs.~\cite{CMS:2021aly,ATLAS:2022wec}.

\section{Machine learning for a probe of EFT operators in the associated production of top quarks with a Z boson at the CMS experiment}

The search for physics beyond the SM is crucial to increase our knowledge of the fundamental mechanisms of particle physics.
For this purpose, effective field theories~(EFTs) have proven themselves as a promising tool.
Their general concept is to extend the known Lagrange function $\L$ with additional operators $\O$ such that
\begin{equation*}
	\L = \L_{\text{SM}} + \sum_{d>4}^{N}\sum_{i} \frac{c_{i}^{(d)}}{\Lambda^{d-4}} \O_i^{(d)}.
\end{equation*}

In this equation, $d$ refers to the dimension of the operators, $N$ denotes the maximum dimension of additional operators considered and $\Lambda$ denotes the energy scale for new physics.
The parameters $c_i$ are known as the Wilson coefficients and indicate the contribution of the corresponding operator to the extension of the SM Lagrangian $\L_{\text{SM}}$.

The CMS analysis described in Ref.~\cite{CMS:2021aly} probes these coefficients in the context of the production of top  quarks in association with Z bosons.
To this end, the analysis considers three signal processes that are sensitive to a different variety of operators: the production of a top quark-antiquark pair in association with a Z boson (\ttZ) as well as the production of a single top quark in association with a Z boson and a W boson (\tZW) or an additional jet (\tZq).
The analysis defines two signal regions based on the number of leptons in the final state as well as additional control regions designed to control the dominant background processes during the estimation of the Wilson coefficients relevant for this analysis.
ML algorithms are employed in the region with three leptons, which is the focus of the following paragraphs.\\

\begin{figure}
	\begin{center}
		\includegraphics[width=\textwidth]{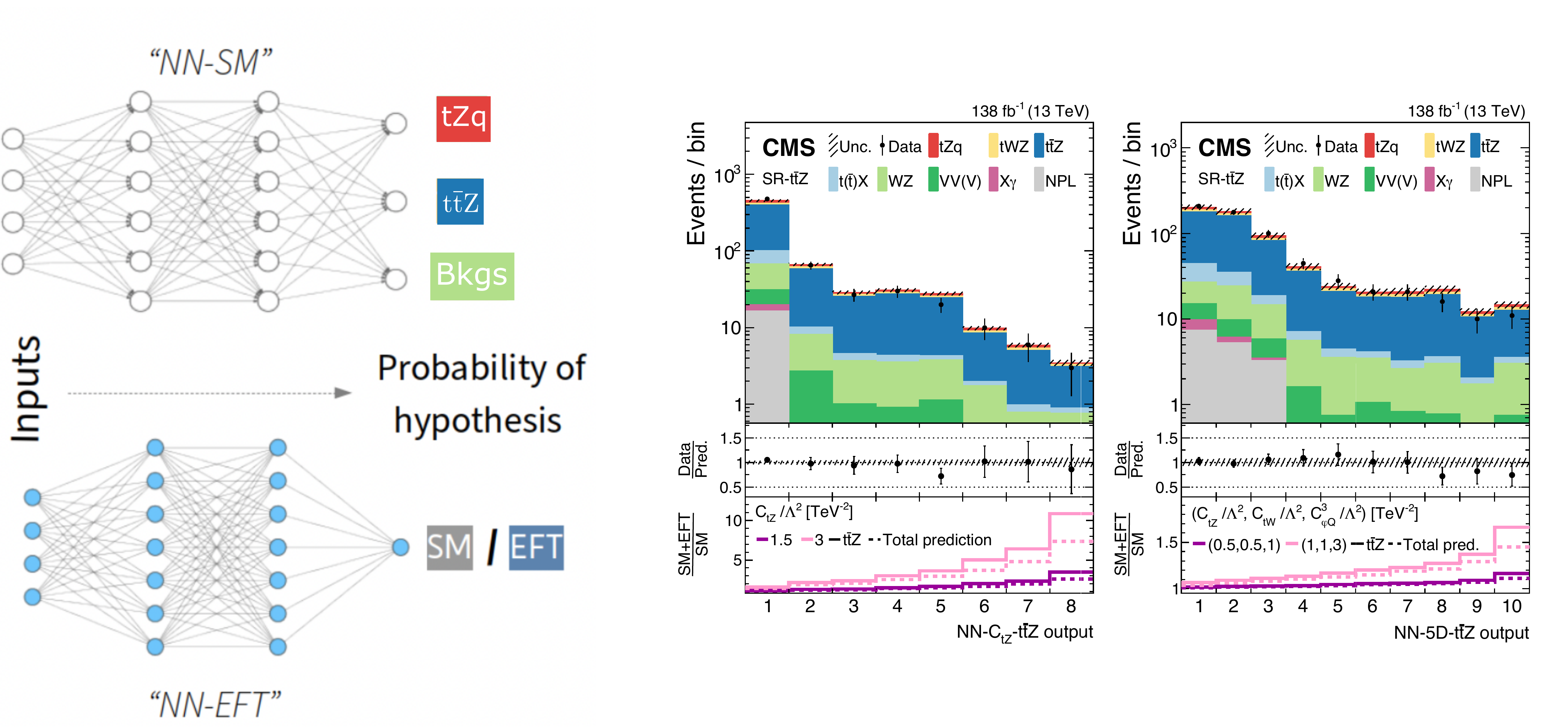}
	\end{center}
	\caption{Events are first categorized according to three classes within the SM (top left).
		Events classified as either \tZq or \ttZ are then considered for a binary classification to increase the sensitivity to EFT contributions (bottom left).
		The resulting distributions are expected to be sensitive to different values of the corresponding Wilson coefficients~(right).
		Figure adapted from Ref.~\cite{CMS:2021aly}.	
	}
	\label{fig:CMS_EFT:eft_classifiers}
\end{figure}

The ML application in this analysis can be divided into two parts.
First, events are categorized using a multi-classification approach, which is based on kinematic variables, a b tagging discriminant as well as other more complex variables.
The architecture of this classifier is summarized in Fig.\ref{fig:CMS_EFT:eft_classifiers}.
It employs three fully-connected hidden layers with 100 neurons in each layer, which are activated with the ReLU function.
The loss function during the training consists of the sum of the categorical cross-entropy function and L2 regularization.
To mitigate possible over-training effects, dropout and early stopping are employed during training, which is based on simulated events corresponding to the dominant physics processes in the analysis phase space.
The target of this classification is to attribute these events according to three categories.
The first two categories correspond to the \tZq and the \ttZ signal processes, while all other events are categorized as background processes.
The \tZW signal process is not specifically considered in the categorization task due to its small contribution to the phase space and the significant kinematic similarities with the other signal processes.
The output values of the final layer are activated using the softmax function, which assigns each event a probabilistic value for each of these classes.
Events are attributed to the class with the largest activation value in the final layer.
This approach yields excellent control over the SM processes, which is consequently used to increase the sensitivity to the EFT parameters under scrutiny.\\

The second part of the ML application in this analysis targets possible EFT contributions to the signal processes under consideration of interference effects with the SM and other EFT operators.
To this end, different sets of binary classifiers are trained with events categorized as either \ttZ or \tZq by the previously-described multi-classification, respectively.
Each set consists of classifiers designed to separate contributions under the SM hypothesis from contributions of either specific EFT operators or their superposition, resulting in a total of eight binary classifiers.
A schematic representation of these classifiers is shown in Fig.~\ref{fig:CMS_EFT:eft_classifiers}.
Each classifier is based on a dedicated set of input features which are propagated through an architecture with three hidden layers consisting of 50 rectified linear units, respectively.
The final output layer employs the sigmoid activation function.
The loss function of these classifiers consists of the binary cross-entropy function as well as a L2 regularization term.

During the training, events are weighted according to the probability density obtained for a wide range of values for the Wilson coefficients corresponding to the most dominant EFT operators.
This facilitates a training under consideration of different interference effects without a bias to preexisting results.
An example for the expected sensitivity to different values of Wilson coefficients is shown in Fig.~\ref{fig:CMS_EFT:eft_classifiers}, which indicates increasing separation power for larger contributions of EFT operators.
The authors of Ref.~\cite{CMS:2021aly} show that this strategy can increase the sensitivity to different Wilson coefficients by up to $70\,\%$, indicating the importance of ML in this analysis.

\section{Machine learning in a measurement of charge asymmetry with the ATLAS experiment}

\begin{figure}[t]
	\begin{center}
		\includegraphics[width=\textwidth,page=1]{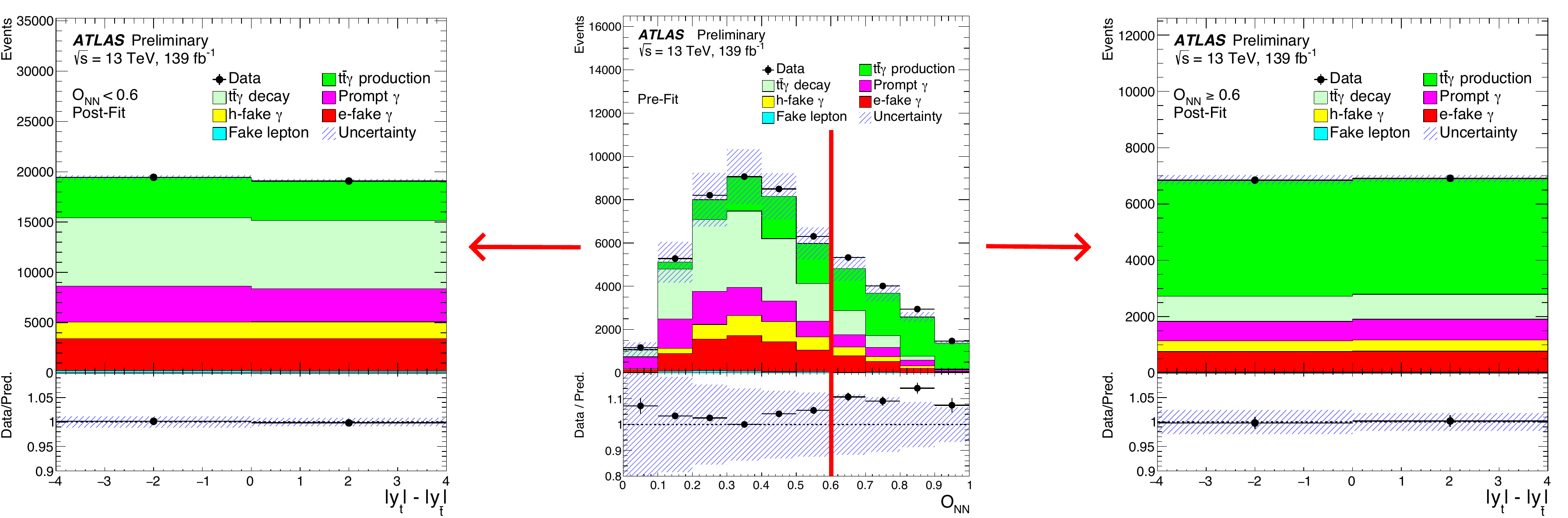}
	\end{center}
	\caption{Shown is the distribution of output values of the binary classifier for simulated events as well as observed data (middle panel).
	Events are divided into two regions based on this distribution (red line), and subsequently used in the unfolding procedure (panels on left and right).
	Figure adapted from Ref.~\cite{ATLAS:2022wec}.
	}
	\label{fig:ATLAS:categorization}
\end{figure}

Another promising avenue in the search for physics beyond the SM is to perform high-precision measurements in the context of the SM itself.
In these approaches, unfolding techniques are powerful and useful tools that aim to revert effects impeding the accuracy of a measurement arising from the detector.
This facilitates a direct comparison of measured quantities with the corresponding theoretical predictions.\\

The analysis presented in Ref.~\cite{ATLAS:2022wec} performs such a measurement for the charge asymmetry between the top quark and antiquark with data obtained with the ATLAS detector.
Such an asymmetry ensues by interference effects at next-to-leading order in quark-initiated production of a \ttbar pair, which is difficult to identify since this production is dominated by gluon-induced processes at the LHC.
Therefore, the analysis requires the presence of an additional photon in the final state to increase the selection efficiency of quark-induced \ttbar events.
The resulting analysis phase space is dominated by background processes, which comprise about $70\,\%$ of the selected events.\\

The analysis relies on a binary classification to further enhance the signal purity.
The model architecture consists of three hidden layers with up to 96 parametric rectified linear units, while the final layer consists of a single node using the softmax activation function.
The binary cross-entropy is used as the loss function during training, which is based on simulated events and kinematic properties of the final state particles as well as additional information about the b tagging and the event shape.
Apart from a separation of the training data into training, validation and test data sets, the analysis employs 5-fold cross validation, which enables an optimal usage of the available events.
The output values of this neural network is subsequently used to define a background-enriched and a signal-enriched region in the phase space, which is illustrated in Fig.~\ref{fig:ATLAS:categorization}.
This reduces the background contamination in the signal region to approximately $45\,\%$, thus greatly increasing the sensitivity of this analysis.

\section{Future applications}

Machine learning in the top quark sector is a fast evolving field in both the ATLAS and CMS Collaborations.
On the one hand, this involves augmenting existing applications of ML with novel architectures.
One example of this approach is presented in Ref.~\cite{Qu:2022mxj}, where the existing b tagging algorithm used within CMS was enhanced by a Transformer architecture.
This increases the selection efficiency of b jets, which is expected to have a direct impact on the reconstruction efficiency of top quarks.

Apart from researching novel architectures, understanding and improving the training samples is equally crucial to improve future applications.
An example for this type of research is presented in Ref.~\cite{ATLAS:2022qby}.
The authors investigated a subset of architectures that showed good identification power for top quarks in the study presented in Ref.~\cite{Kasieczka:2019dbj}, which was based on a fast detector simulation.
In Ref.~\cite{ATLAS:2022qby}, these models were trained with a full detector simulation, which is expected to increase the precision of the simulations used during the training.
The authors show that the performance of these architectures can depend significantly on the precision of the detector simulation, indicating the need to consider such effects when constructing and training a ML algorithm.
The data set used in this study is publicly available in CERN OpenData.

On the other hand, novel fields for ML applications are under investigation.
One of the most challenging aspects in top-quark related analyses is currently the estimation of systematic uncertainties, which is a critical aspect in many measurements.
Due to the high complexity of the underlying effects, the estimation of these uncertainties often requires dedicated simulations, which are computationally expensive.
These Monte Carlo samples are therefore produced with a limited amount of events, which can impede the statistical power of the estimation depending on the phase space used in a given analysis.

One possible solution for this problem could be the DCTR approach, which is presented in Ref.~\cite{Andreassen:2019nnm}.
The authors show that a binary classification to separate simulations with different parameter values of the underlying generator can be used to reweigh the events  from one set into the other.
This would potentially enable a continuous reweighing procedure based on the full phase-space information, thus increasing the interpretability of the uncertainty estimates and reducing the computational expenses.
The applicability of this method within the collaborations is currently under investigation.

\Acknowledgements
Special thanks to my colleagues in the ATLAS and CMS Collaborations for providing material and reviewing this work.

\nocite{*}

\newpage
\printbibliography
 
\end{document}